\documentclass[runningheads]{llncs}

\usepackage{graphicx}
\usepackage{natbib}

\titlerunning{Socially Interactive Agents for Robotic Neurorehabilitation Training}
\authorrunning{R. Arora et al.}

\def\Authors{Rhythm Arora\,$^{1,*}$, Pooja Prajod\,$^{2,\dagger}$, Matteo Lavit Nicora\,$^{3,4,\dagger}$, Daniele Panzeri\,$^{5}$, Giovanni Tauro\,$^{3,4}$, Rocco Vertechy\,$^{4}$, Matteo Malosio\,$^{3}$, Elisabeth André\,$^{2}$, and Patrick Gebhard\,$^{1}$}
\def\Address{
$^{1}$German Research Center for Artificial Intelligence, Saarbrücken, Germany\\
$^{2}$Human-Centered Artificial Intelligence, Augsburg University, Augsburg, Germany\\
$^{3}$National Research Council of Italy, Lecco, Italy\\
$^{4}$Industrial Engineering Department, University of Bologna, Bologna, Italy\\
$^{5}$Scientific Institute IRCCS E. Medea, Bosisio Parini, Lecco, Italy\\
$^{\dagger}$These authors contributed equally to this work.\\
$^{*}$Corresponding Author, rhythmarora0406@gmail.com
}

\author{\Authors}
\institute{\Address}

\begin{document}

\title{Socially Interactive Agents for Robotic Neurorehabilitation Training: Conceptualization and Proof-of-concept Study}



\maketitle

\begin{abstract}
Individuals with diverse motor abilities often benefit from intensive and specialized rehabilitation therapies aimed at enhancing their functional recovery. Nevertheless, the challenge lies in the restricted availability of neurorehabilitation professionals, hindering the effective delivery of the necessary level of care. Robotic devices hold great potential in reducing the dependence on medical personnel during therapy but, at the same time, they generally lack the crucial human interaction and motivation that traditional in-person sessions provide. To bridge this gap, we introduce an AI-based system aimed at delivering personalized, out-of-hospital assistance during neurorehabilitation training. This system includes a rehabilitation training device, affective signal classification models, training exercises, and a socially interactive agent as the user interface.
With the assistance of a professional,  the envisioned system is designed to be tailored to accommodate the unique rehabilitation requirements of an individual patient. Conceptually, after a preliminary setup and instruction phase, the patient is equipped to continue their rehabilitation regimen autonomously in the comfort of their home, facilitated by a socially interactive agent functioning as a virtual coaching assistant. Our approach involves the integration of an interactive socially-aware virtual agent into a neurorehabilitation robotic framework, with the primary objective of recreating the social aspects inherent to in-person rehabilitation sessions. We also conducted a feasibility study to test the framework with healthy patients. The results of our preliminary investigation indicate that participants demonstrated a propensity to adapt to the system. Notably, the presence of the interactive agent during the proposed exercises did not act as a source of distraction; instead, it positively impacted users' engagement.

\end{abstract}

\section{Introduction}
\label{sec:intro}

Neurorehabilitation is a widely used medical practice that aims to aid recovery from a nervous system injury. Its purpose is to maximize and maintain the patient's motor control while trying to restore motor functions in people with neurological impairments. Given the constant growth and aging of the world population, the number of patients affected by neuromotor disorders that seek the attention of professionals for their rehabilitation therapy is constantly increasing~\citep{Crocker2013}. However, due to a lack of medical personnel, it is impossible to provide the intense training that would be needed for an effective recovery of the patient's capabilities, therefore hindering the actual outcomes of the treatment~\citep{ETDP-6DR4-D617-VMVF}.

This situation is both harmful for the patients and constitutes a relevant burden on society and the healthcare system~\citep{Wynford-Thomas20172345}. To address this issue, robot-assisted training has been widely investigated as an effective neurorehabilitation approach that helps augment physical therapy and facilitates motor recovery. 
According to the literature~\citep{Maciejasz-2014, Zhang-2017, Qassim-2020}, such approaches can help therapists save time while reproducing accurate and repetitive motions for the patients' high-intensity training. 
In particular, upper-limb robotic rehabilitation is one of the fastest-growing areas in modern neurorehabilitation. 
Leveraging the capabilities of these robotic systems, their application in domestic environments would represent a plausible solution to the lack of treatment intensity that patients are experiencing nowadays. 
In fact, a system capable of assisting the patient in performing the necessary repetitive motions would relieve a lot of the pressure that is acting on the clinical structures, since the physical presence of medical personnel would be required only for sporadic interventions. 
However, a crucial issue for rehabilitation training is user engagement and motivation~\citep{Blank2014184}, which may be lacking if the rehabilitation system is used without a human medical coach. 
Since the effectiveness of the treatment has been proven to be related to the patient's level of engagement~\citep{Turner-Stokes2015210}, it is important for the envisioned system not only to be able to physically assist the patients but also to understand their affective state and react accordingly. Hence, we believe that introducing a socially-aware interactive virtual agent could represent a promising solution to recreate the social aspects of in-person rehabilitation sessions. 
Moreover, insights from the field of social robotics suggest that the enhancement of a rehabilitation system through affective and social signal processing can augment the personalization of the system and further facilitate neuroplasticity~\citep{NAHUM2013141}. Therefore, a neurorehabilitation training system capable of modeling the patient's state and tuning its behavior depending on both the measured performance deviation index and the inferred mental and physical state could improve the user's engagement and the outcome of the therapy. The performance deviation index, which is inversely proportional to the user's performance, represents the deviation of the actual path followed by the participant from the ideal path.

\begin{figure}[ht]
    \centerline{\includegraphics[width=0.75\textwidth]{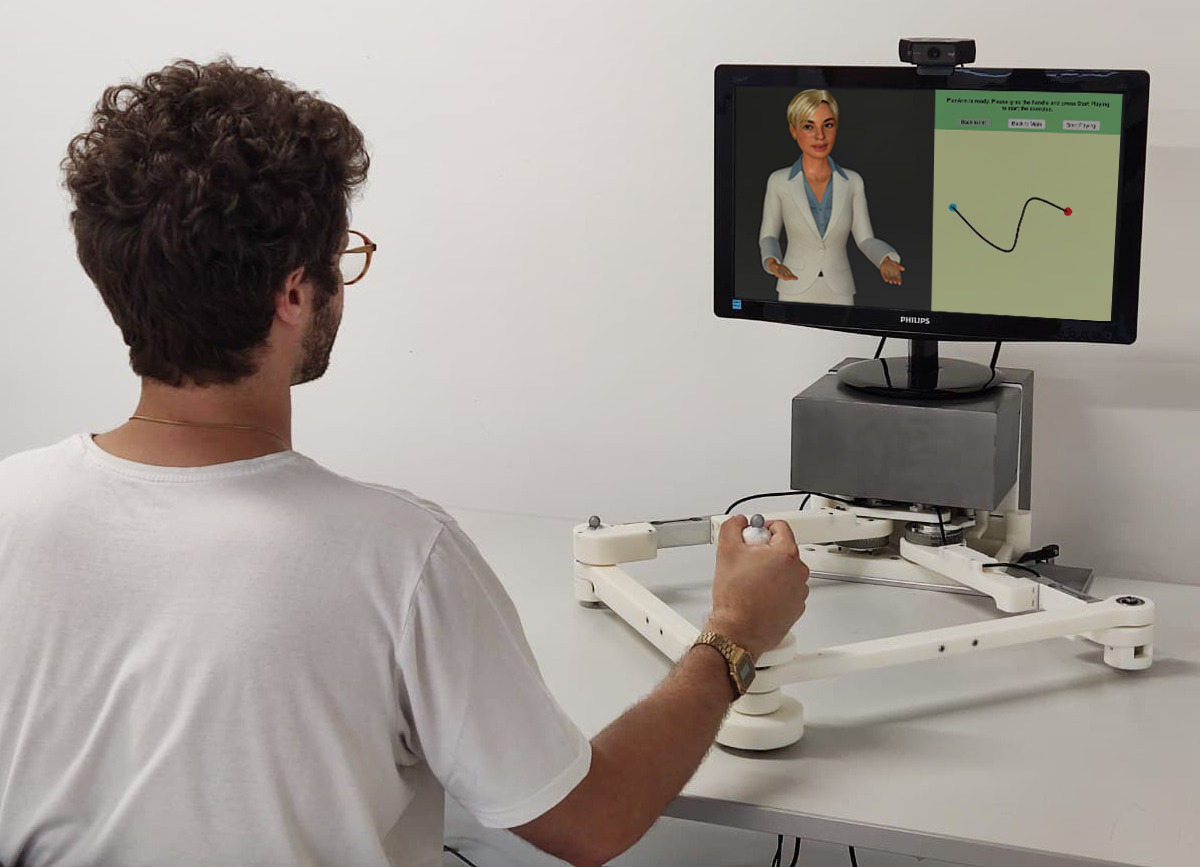}}
    \caption{Empathetic Neurorehabilitation Trainer concept.}
    \label{fig:system_overview}
\end{figure}

This paper investigates the feasibility of the Empathetic Neurorehabilitation Trainer depicted in Figure~\ref{fig:system_overview}, a technology-based upper-limb neurorehabilitation system equipped with social interaction capabilities and composed of 1) a planar rehabilitation device for physical assistance and 2) a socially interactive virtual agent in the role of a supportive coaching assistant. 
In this context, the physiological and behavioral signals of the user are collected, analyzed, and elaborated into attentiveness, stress, and pain information, which are then exploited to tune the rehabilitation session coherently. In this scenario, having a socially interactive agent can help support the patients' engagement and motivation in a flexible and personalized way. It is important to state that professional physiotherapists would still play a fundamental, although less time-consuming, role in this home-based robotic treatment. In fact, the system is envisioned to be used by patients at home only after a training phase. 
During this phase, the system is required to learn directly from the experience of professional physiotherapists how to respond to the needs of the specific patient~\citep{Lequerica2009753}. In particular, the system should be able to understand both the patient's residual physical capabilities, in order to provide a properly tuned level of assistance, and the behavioural patterns that should be elicited by the virtual coach in order to keep the patient engaged in the exercise.

\subsection{Background}
Before going straight into the description of the envisioned Empatic Neurorehabilitation Trainer, an overview of the background knowledge gathered and analysed as a basis for the conceptualization of the system is reported. In particular, an introduction to the 'neurorehabilitation best practices' collected from the experience of professional therapists and supported by literature (Section~\ref{sec:bestpractice}), the fundamental concepts of robotic neurorehabilitation (Section~\ref{sec:robotrehab}), the conventional methods to detect affective states through behavioral and physiological signals interpretation (Section~\ref{sec:ssi}) and the theory behind trustworthiness of virtual agents (Section~\ref{sec:trust}) are presented below.

\subsubsection{Neurorehabilitation best practices}
\label{sec:bestpractice}
In order to gather precious insights on the strategies and struggles that professional neurorehabilitation therapists experience during their daily practice, we decided to not only study the literature on the topic but also perform a series of interviews, performed online through Microsoft Teams as free open discussions. Overall, we were able to collect the point of view of 15 therapists spread over the Italian territory and here a summary of the obtained insights is reported.

The constantly growing employment of technological devices in neurorehabilitation therapy can be explained by the introduced ease in reaching a significant number of movement repetitions in a specific body district (body parts grouped by functionality)~\citep{Panzeri1} and in obtaining higher patient engagement, fundamental aspects in the motor rehabilitation process. In this regard, one of the interviewees said (translated from Italian): "I like to use technological solutions for the intensity they can provide and for the possibility to perform rehabilitation also in very extreme cases". However, even when relying on robotic solutions, one must be able to balance movement repetition, often leading to boredom, with variability and incentive to enhance the patient's attention and commitment.

During every therapy session, a therapist assesses the actual level of attention, commitment, engagement, stress, and pain currently experienced by the patient to deliver the correct amount of exercise and to avoid the risk of too easy or too difficult tasks that may lead to a decay in interest or even a feeling of frustration~\citep{Panzeri2, Panzeri3, Panzeri4}. In order to reach this goal, the therapist relies on the activity scores and the patient's behavior. For example, if the patient cannot achieve a particular performance deviation index , the selected activity is likely too difficult. On the other hand, if the patient can perform the exercise but, after some time, becomes very talkative and less performing, it is likely that a decay in interest is occurring. In such cases, the therapist must give feedback and, if needed, support when the activities are too difficult or change the exercise when the attention starts to decrease. Furthermore, considering patients affected by neurological impairments, attention problems are frequent. Recording the period to which the attention lasts can be valuable information to provide a correct dosage of exercise. Lastly, the management of neurological disorders can be considerably different between adults and children. In both cases, understanding when a pause or a change of exercise is necessary is crucial. In this regard, one of the interviewees said (translated from Italian): "The currently available robotic devices often require the patient to adapt to them instead of the opposite. An automatic tuning of the exercise duration and difficulty to the needs of the patients would be game-changing.". Moreover, considering adults, one can count on their responsibility to train towards an improvement, even if the activity could lead to boredom. On the other hand, children may not behave in the same way and, to augment their engagement, it is crucial to introduce gaming aspects to the exercise.

\subsubsection{Robotic neurorehabilitation}
\label{sec:robotrehab}

In the last two decades, robotic devices for neurorehabilitation have been widely investigated to offer a valid alternative to conventional therapy and to fill the constantly growing gap between supply and demand, both for upper-limb and lower-limb rehabilitation~\citep{Maciejasz-2014,Zhang-2017,Qassim-2020}. These robots are generally well tolerated by patients and have been found to be an effective adjunct to therapy in individuals with motor impairments~\citep{Zhang2013, Lo20173049, Carpino2018}.

A number of benefits can be found to support the decision to introduce this kind of systems in the available set of neurorehabilitation treatments. Rehabilitation robots are able to objectively measure the amount and type of assistance provided during therapy and to actively track changes in motor functions that occur during the progression of the therapy. Moreover, these devices are designed to augment the clinicians’ toolbox and to allow them to deliver meaningful restorative therapy to more patients by enhancing both productivity and effectiveness as they try to facilitate the individual’s recovery~\citep{krebs2000increasing}. Three main aspects can be said to characterize robotic rehabilitation: 

\begin{itemize}
    \item \textbf{Repeatability}. Reliable repetition of the process or exercise without any physical effort by the therapist.
    \item \textbf{Measurability}. Exact, quantitative and objective measurements thanks to the sensors mounted on the device.
    \item \textbf{Intensity}. Administration of intensive rehabilitation tasks that can also be autonomously performed by the patient.
\end{itemize}

Given the above benefits, the medical community has increasingly recognized the importance of robotic rehabilitation. However, most of the robots available on the market are relatively expensive, making them accessible only to large rehabilitation centers. A new trend of cost-effective domiciliary devices would therefore unleash the spread of rehabilitation robots within the home environment for many patients. This not only requires the hardware components to be designed on the basis of cost-optimization strategies, but also the software modules need to be developed with a user-centered approach, in order to make the whole system intuitive and usable by anyone without the need for special training or expert personnel.

The first goal of a neurorehabilitation robot control algorithm is the ability to elicit neuroplasticity and enhance the patient's motor recovery. To make this possible, it is crucial that the assistance provided by the device is not too low in order to allow the patient to complete the task and to avoid frustration, but also not too high, thereby ensuring that the patient actively participates in the task with no risk of slacking~\citep{erdogan2012slacking}. Also, it is important that the device does not perfectly correct the motion initiated by the user. In fact, a certain amount of error has been proved to be useful in stimulating neuroplasticity given that the patient has to put focus and effort on the task in order to correct the motion autonomously as much as possible~\citep{takagi2018reduced}. Thus, the capability of a robot to actively and automatically adapt the level of assistance according to the skills and the performances of the patient is one of its most important features~\citep{Marchal-Crespo-2009, Meng-2015}.

Furthermore, the level of provided assistance should not be defined only on the basis of the kinematic performances of the patient. In this regard, the evaluation of social, physiological and psychological aspects provides a more fine-grained assessment of the patient's state, useful to achieve a better tuning of the behaviour of the system~\citep{Novak-2011, Malosio-2016}. For instance, as mentioned in Section~\ref{sec:bestpractice}, a patient that starts feeling bored will be less engaged on the task with the risk of reducing the effectiveness of the exercise. A system capable of detecting this state could, instead, render the task more challenging, for instance by reducing the level of provided assistance, to bring the patient's focus back on the exercise.

\subsubsection{Affective Signal Interpretation}
\label{sec:ssi}
During the training sessions, the affective signals collected from the patients can be used to infer useful information about their experience. Home-based healthcare systems frequently leverage a diverse range of affective signals~\citep{majumder2017smart, philip2021internet, wang2021unobtrusive}. In this section, we provide a brief description of three affective states integral to neurorehabilitation, along with an overview of typical modalities utilized for inferring these states.

\begin{itemize}
    \item \textbf{Attention}: Motivation and attention serve as crucial modulators of neuroplasticity, influencing the outcomes of rehabilitation therapy~\citep{Cramer-2011}. Distractions, stemming from factors like boredom or lack of motivation, can disrupt the user's engagement during training sessions. Hence, the user's attention level becomes a pivotal input for the agent's motivational strategy in neurorehabilitation. While previous studies in various domains have demonstrated the prediction of attention through physiological signals such as EEG~\citep{c11, souza2021attention}, these methods require proper sensor placement and additional user training on sensor usage. A more practical alternative lies in camera-based solutions, which capitalize on a common behavioral cue associated with distraction — looking away from the task. Research in other domains~\citep{c12, smith2003determining, prajod2023gaze} has indicated that facial and body pose features, including gaze direction, head orientation, and body posture, can effectively detect loss of attention. Inferring attention from such features is contingent on the setup (e.g., screen position), and detection models need to be appropriately calibrated. Nonetheless, this approach presents a cost-effective and unobtrusive solution when compared to sensors like EEG.

    \item \textbf{Pain}: Research on the occurrence of pain within the neurorehabilitation population and the consequent necessity for medical interventions has been extensively explored in works dedicated to neurorehabilitation~\citep{benrud2000chronic, castelnuovo2016psychological}. In the realm of healthcare applications, numerous systems employ image or video-based automatic pain detection~\citep{c3, c4}. These approaches typically entail the identification of pain based on facial expressions captured by a frontal camera. Some works~\citep{lopez2018continuous, werner2014automatic} have also delved into the utilization of physiological signals such as ECG, EDA, etc., for pain recognition. Despite recent strides in affective computing toward automatic pain detection, the available datasets remain limited in size, often necessitating techniques like transfer learning to address this constraint~\citep{c6, prajod2021deep, prajod2022using}.

    \item \textbf{Stress}: Detecting stress becomes crucial, especially with the introduction of gamification elements in the training session, where the patient may experience stress, particularly if the exercise surpasses their current skill level. Extensive research has explored diverse modalities for stress detection, encompassing physiological signals, speech, gestures, and contextual behavioral patterns~\citep{koceska2021review, larradet2020toward, giannakakis2019review, heimerl2023fordigitstress}. Physiological signals, including ECG, BVP, EDA, and respiration, have demonstrated high efficacy in stress detection~\citep{gedam2021review, prajod2024stressor, smets2018into}. Audio or speech analysis is another prevalent modality for automatic stress recognition~\citep{dillon2022voice, lefter2015recognizing}. However, this approach typically involves substantial verbal interaction with the agent, a scenario not anticipated during neurorehabilitation exercises. Contextual behaviors, such as keystrokes and specific gestures, are often tailored to specific use cases and may not be directly applicable in the context of neurorehabilitation.

\end{itemize}

\subsubsection{Warmth and competence}
\label{sec:trust}

In the pursuit of developing socially interactive agents for neurorehabilitation, our aim is to not only create effective agents but also ensure that they are perceived as warm and competent. These perceptions of warmth and competence hold substantial importance as they underpin the establishment of trust and user engagement, both of which are integral to the success of our approach. In these terms, several key factors emerge as critical determinants of success. Anthropomorphism, which involves attributing human-like qualities to non-human entities, plays a fundamental role in cultivating a sense of warmth in social agents~\citep{nass2000machines,Lee2006Physically}. Users tend to respond more positively to agents that exhibit anthropomorphic traits, perceiving them as approachable and friendly, and this perception of warmth significantly contributes to users' overall experiences and their willingness to cooperate with the technology~\citep{Prajod2019Effect}. 

Competence, another essential factor, has been identified in psychology research as a critical determinant of trust~\citep{Hancock2004Deception,Bickmore2009Using}. An agent's competence, reflecting its capabilities and effectiveness, directly influences the trust users place in it~\citep{Hancock2004Deception,Bickmore2009Using}. Users are more inclined to trust and cooperate with agents they perceive as competent in assisting them with their rehabilitation tasks. Both warmth and competence exert substantial influence on user engagement, as research demonstrates that users are more engaged and motivated to interact with agents perceived as both warm and competent~\citep{nass2000machines}. This heightened engagement is of paramount significance in neurorehabilitation, as it bolsters users' commitment to therapy and increases the likelihood of positive outcomes~\citep{nass2000machines}. 

To enhance the warmth and competence of social agents in the context of neurorehabilitation, careful consideration must be given to anthropomorphic design elements, including the incorporation of human-like features, gestures, and verbal and non-verbal communication styles, all of which may elicit feelings of warmth and trust in users~\citep{nass2000machines,Lee2006Physically}. Our research underscores the importance of integrating anthropomorphic design, effective communication strategies, and a robust knowledge base in crafting agents that not only provide effective assistance to patients but also cultivate trust, foster engagement, and contribute to positive rehabilitation outcomes. It is evident that further investigation is warranted to delve deeper into the nuances of how warmth and competence perception influence user engagement and, ultimately, the outcomes of neurorehabilitation interventions.

\subsection{Related works on socially interactive agents as medical coaches}

In the realm of neurorehabilitation, the use of socially interactive agents as medical coaches has garnered increasing attention due to their potential to enhance patient engagement and therapeutic outcomes. This section provides an overview of related works in this domain, shedding light on the role of virtual coaches and their impact on patient motivation and progress.

Within the domain of virtual coaching and interactive systems for medical applications, several notable initiatives have paved the way for the development and implementation of socially interactive agents in various healthcare contexts. \cite{Bickmore-et-al-05} introduced the Fit Track system, featuring the relational agent Laura, who serves as an exercise advisor. Laura engages with patients, motivating them to participate in physical activities and thereby fostering their rehabilitation progress. This system stands as an early exemplar of virtual coaching in the medical field. The SenseEmotion project~\citep{velana2017senseemotion} explored pain management strategies among the elderly, employing an avatar for crisis interventions to facilitate reassuring dialogues and support for older adults. This initiative highlighted the potential of avatars in the context of pain management. \cite{Schneeberger-et-al-21} delved into stress management using virtual characters in simulated job interview scenarios. Their research aimed to understand the various interaction strategies these virtual characters could employ to induce stress in participants, offering valuable insights into human-agent interaction and applications in training and preparatory systems. \cite{neumann2023impact} investigated the impact of virtual social support on physiological pain responses within a virtual reality environment, showcasing the potential for virtual characters to provide emotional support and reduce physiological pain responses. Additionally, \cite{giraud:et:al:2021} proposed a tangible and virtual interactive system to train children with Autism Spectrum Condition in joint actions, demonstrating the broader potential of socially interactive agents in training social and motor skills relevant to neurorehabilitation. Nadine, a Digital Human Cardiac Coach, was developed to support heart patients throughout their cardiac health journey.

In this study, we strategically choose to build upon the Gloria biofeedback training system, a well-established platform developed by our Affective Computing group. By leveraging Gloria's robust framework, we ensure continuity in technological advancement and capitalize on its pre-existing integration capabilities with our current methodologies. This deliberate choice is guided by the system's demonstrated success~\citep{Schneeberger-et-al-21}, providing a solid foundation for adaptation of its behavior and training strategies to specifically address the unique requirements of neurorehabilitation . Our primary objective is to conceptualize a comprehensive system that seamlessly integrates a motivating virtual coach with a robotic rehabilitation device to optimize patient engagement and enhance rehabilitation outcomes. Through this exploration, we aim to address the unique challenges and opportunities presented by neurorehabilitation, emphasizing the potential of socially interactive agents as valuable allies in the journey toward patient recovery. By synthesizing these related works, we draw inspiration from the successes and insights offered by virtual coaches in various medical domains, applying them to the specialized context of neurorehabilitation and striving to empower patients and facilitate their path to recovery.

\section{Materials and methods}
\label{sec:material}

\subsection{Concept}
\label{sec:concepts}
\begin{figure}[thpb]
    \centering
    \includegraphics[width=0.99\textwidth]{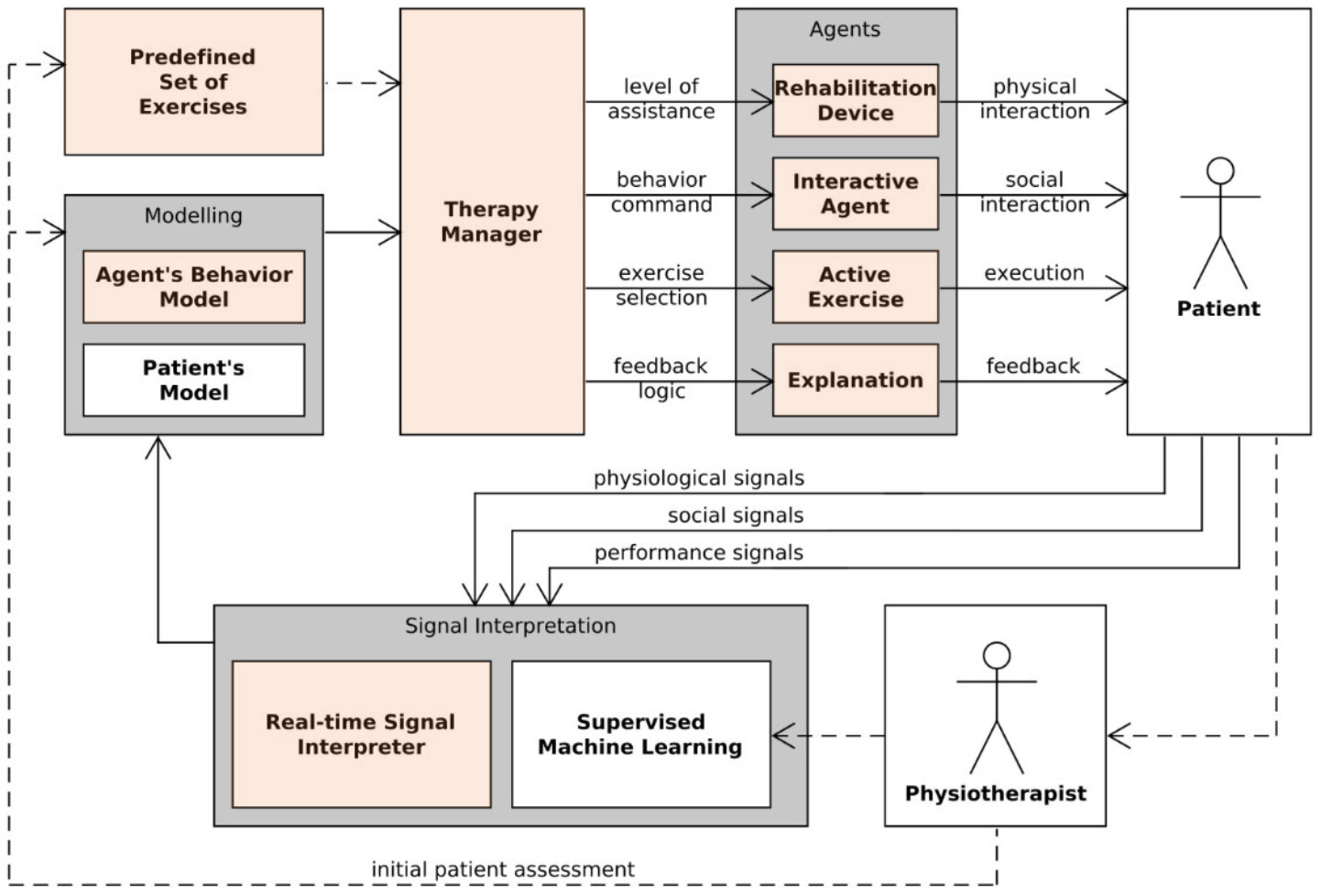}
    \vspace{0.5em}
    \caption{Human-in-the-loop Empathetic Neurorehabilitation Trainer concept}
    \label{fig:deployment}
\end{figure}

The envisioned Empathetic Neurorehabilitation Trainer architecture is shown in Figure~\ref{fig:deployment}. The system is equipped with a robotic \textit{Rehabilitation Device} and a virtual socially \textit{Interactive Agent}. Both device and agent can adapt their behavior based on the \textit{Patient}'s performance, as in most assistance-as-needed paradigms, and it also takes into account the patient's affective state. Moreover, these two entities, together with the specific task to be carried out and the feedback media chosen to provide the patient with an \textit{Explanation} regarding the \textit{Active Exercise}, are intended to work as a single entity, actively collaborating to improve the rehabilitation session outcomes further.

Always with reference to Figure~\ref{fig:deployment}, all software and hardware components make up a closed-loop architecture where the monitoring of a set of heterogeneous parameters is introduced. In fact, during the execution of the task, the robotic device is in charge of collecting data regarding the kinematics of the patient's movement (e.g., position, speed), and a wearable device is used to extract various physiological signals (e.g., ECG, EDA). At the same time, a camera captures the patient's upper body for behavioral signal interpretation purposes. These raw data represent the input for a \textit{Signal Interpretation} module, responsible for providing a series of higher-level quantities such as patient's performance, attentiveness, and stress. As depicted, the \textit{Physiotherapist} still has a central role in the proposed approach. In fact, professional expertise is required for the patient's initial assessment (e.g., residual mobility, attention span), used to define the backbone of both a \textit{Patient's Model} and an \textit{Agent's Behavior Model} and to prepare a \textit{Predefined Set of Exercises}. Moreover, a \textit{Supervised Machine Learning} module is employed to learn from the physiotherapist how to optimally balance the target execution performance for the exercise and the social experience for the specific patient. Also, both challenging and entertaining portions of the session must be included to maximize the patient's attention. Closing the loop, a \textit{Therapy Manager} actively exploits the inferred information to decide how the behavior of the socially interactive agent should be changed, which explanations should be given, and which exercise and difficulty level should be activated to optimize the therapy experience and effectiveness.

Considering Figure~\ref{fig:deployment}, the highlighted modules are the ones that were not only conceptualized but completely realized and tested with a small set of volunteers to verify the correct functioning of all the components. A detailed description of those modules is provided in the following sections.

\subsection{Robot Control}
\begin{figure}[h]
    \centering
    \includegraphics[width=0.6\textwidth]{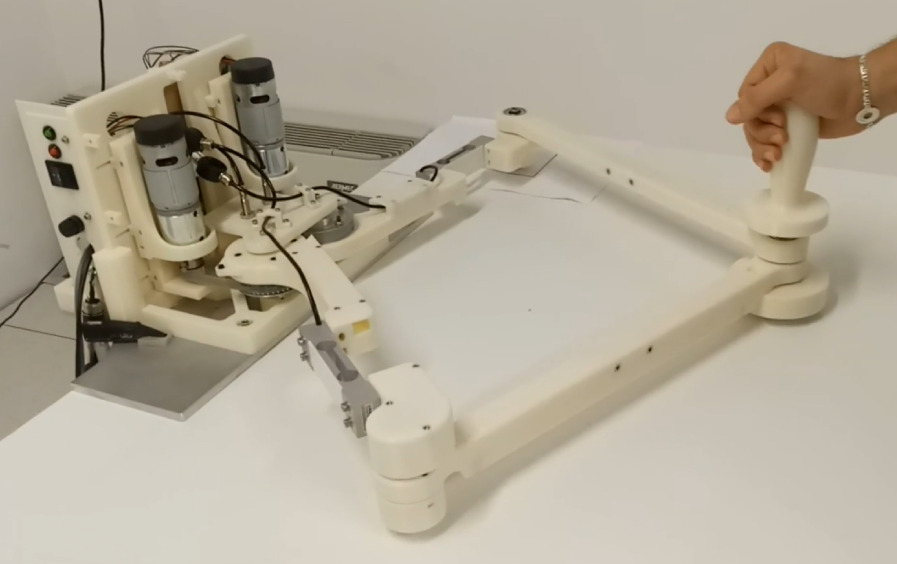}
    \caption{The PlanArm2 prototype.}
    \label{fig:planarm}
\end{figure}

As already mentioned, the backbone of the presented architecture is based on the Robot Operating System (ROS)~\citep{quigley2009ros}. In particular, the whole robot control software has been realized using ROS Noetic on an Ubuntu 20.04 machine, leveraging the built-in functionalities of \textit{ros\_control}~\citep{chitta2017ros_control} and \textit{MoveIt!}~\citep{coleman2014reducing}. Thanks to this approach, the whole architecture is independent of the choice of specific robotic device selected for the rehabilitation practice. However, in order to validate its correct functioning, the system is tested on a prototypical device called PlanArm2~\citep{yamine2020planar}. This 2-DOF planar upper-limb rehabilitation robot, depicted in Figure~\ref{fig:planarm} is selected because of its affordable and compact design, perfectly suited for home-based therapy applications. In simplified terms, the implemented robot control system waits for a command containing the trajectory to be executed by the patient as part of the exercise. The latter is defined by the therapist using a dedicated Graphical User Interface (GUI), presented in Section~\ref{sec:gui}, and sent to the active controller. As the patient starts moving the robot handle along the predefined trajectory, the controller monitors the current handle position with the relative ideal position on the trajectory and generates an assistive restoring force if this error overcomes a certain threshold. During the exercise, the actual position of the robot handle is also communicated back to the GUI both for generating a visual feedback for the patient and for monitoring the execution performance.

\subsection{Graphical User Interface}
\label{sec:gui}
\begin{figure}[h]
    \centering
    \includegraphics[width=0.99\textwidth]{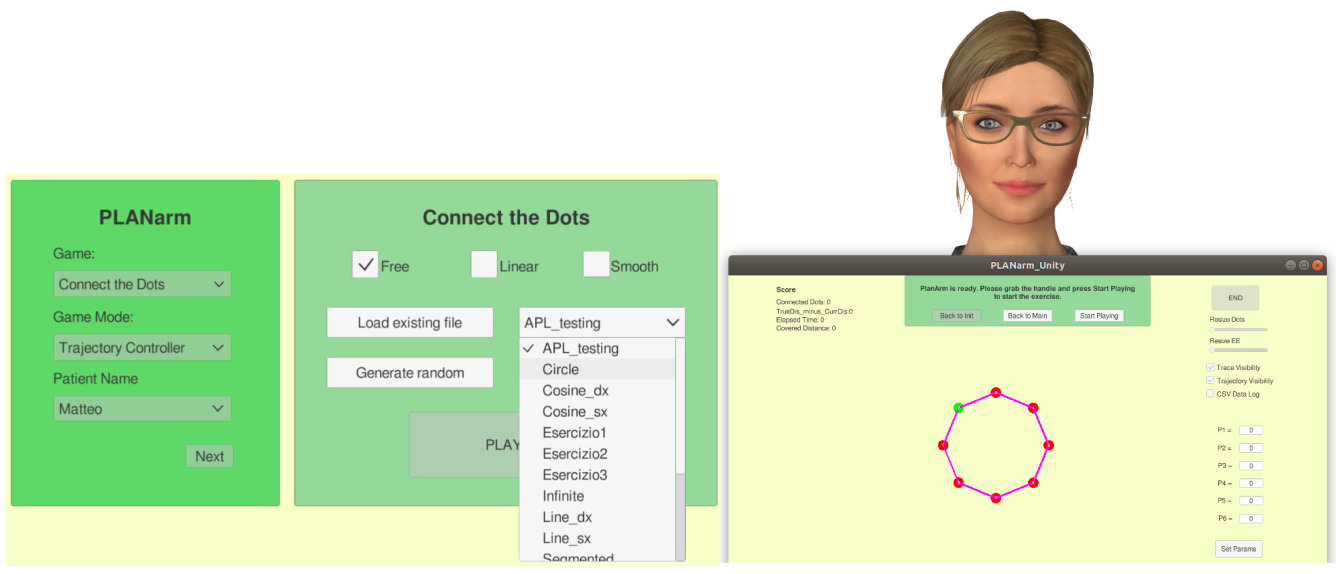}
    \caption{On the left a screenshot of the game and assistance selection page. On the right a screenshot of the visual feedback provided to the patient together with the virtual character.}
    \label{fig:gui}
\end{figure}

In order to render the whole system more user-friendly both for the therapist and for the patient, a graphical user interface was developed using Unity3D~\citep{haas2014history}. With reference to Figure~\ref{fig:gui}, the GUI allows the therapist to select a specific exercise to be performed and a certain setting for the level of assistance provided by the robotic system. On top of that, the application provides the patient with simple visual feedback of the exercise in the form of a game. The developed application, as already mentioned, directly communicates with the robot control system through ROS using the capabilities provided by~\cite{UnityRoboticsHub}. Moreover, as depicted in Figure~\ref{fig:gui}, the application is built to visualize a virtual coach. The GUI, as described earlier, offers patients three distinct types of exercises: a circle (refer to Figure~\ref{fig:gui}),  a straight line, and an infinity. The objective for users in these exercises is to closely follow these paths, with their adherence to the line influencing their scoring (refer to the right hand side of the  Figure~\ref{fig:gui}). Priority is given to the user's Trajectory Error (the deviation of the actual path followed by the participant from the ideal path), followed by the distance traveled, the time it took the user to complete the task, and also the tracking error value, which informs the user if they moved away from the given ideal trajectory. These key metrics are meticulously recorded from the onset of each training game and are subsequently transmitted in real-time using UDP Sockets to the visual scene maker (VSM)~\citep{Gebhard-et-al-12} at the end of each training game. Each participant was required to complete three sessions of the selected exercise. The total duration of the training for each individual was approximately 15 minutes. At the conclusion of the training, the participant's performance deviation index  was compared across all three sessions. Here, in the VSM, the interactive agent verbally communicates the user's performance deviation index results (see section~\ref{sec:results}, for more details about how performance deviation index is computed).

\subsection{Machine Learning Models}
Various affective and physiological signals can be used to analyze the mental and physical states of the user. We trained machine learning models to detect distraction, pain, and stress. Distraction and pain are detected from the images captured by a frontal camera. Stress is detected using the ECG signals recorded through a chest band worn by the user. The detection results of these states serve as important inputs to the agent's motivational strategy.

\subsubsection{Attention Detection}
The patient may become distracted during the training session due to boredom or lack of motivation. In neurorehabilitation, the patient's level of attention is a crucial input for the agent's motivational strategy. Distraction or lack of attention is a key state to detect in this context. We say a user is distracted when they are paying attention to the surroundings rather than the training exercise screen. We can redefine the problem as a use case of gaze estimation, where the user's gaze on the screen is considered attentive and anywhere else is considered distracted. 

We train a deep neural network for discerning attention to screen and distraction following the approach proposed by~\cite{prajod2023gaze}. First, we train a VGG16 network for gaze estimation using the ETH-XGaze dataset~\citep{zhang2020eth}. This dataset has high variations in gaze, including extreme head positions. The input images are face-cropped by leveraging the face detection model~\citep{blazeface} provided by MediaPipe. The input images are scaled to the default VGG16 dimensions of 224 $\times$ 224. The network outputs pitch and yaw values corresponding to the gaze direction. 

Next, we need a target dataset for training the network on our specific task. We collected a small dataset (approximately 200 images) of people looking at the screen and distracted (looking away in random directions). Five participants (3 males, and 2 females; aged 18-30 years ) took part in the data collection, of which three wore glasses. High-resolution images (1920 $\times$ 1080 ) were captured using a camera positioned on top of the screen.

Finally, we adopt a transfer learning approach to detect when the user is distracted. We fine-tune the prediction layer of the gaze estimation network using the collected dataset. The prediction layer is adjusted for 2-class classification (screen or away) and uses Softmax activation. Similar to the gaze estimation network, the input images are face-cropped and scaled to 224 $\times$ 224. The fine-tuning is performed using SGD optimizer (learning rate = 0.01) and categorical cross-entropy loss function. We follow leave-one-subject-out (LOSO) evaluation to train the model using data from four participants and reserve the unseen data from one participant for validation. Our model achieves an average accuracy of 84.6\%.

\subsubsection{Pain Detection}
Pain assessment and management are crucial for medical interventions in neurorehabilitation and hence, is a state that needs to be monitored during the therapy session. In this work, we train a deep-learning model that can discern pain and no-pain images from facial expressions captured by the front camera. One major challenge here is that pain datasets are typically small for training deep learning models~\citep{c6, hassan2019automatic, xiang2022imbalanced}. To circumvent this, we follow the transfer learning approach described by~\cite{prajod2021deep}, which involves leveraging features learned for emotion recognition in pain detection. To this end, we train an emotion recognition model using a large dataset called AffectNet dataset~\citep{mollahosseini2017affectnet}. The model uses a VGG16 network to classify an input image as Neutral, Happy, Sad, Surprise, Fear, Anger, Disgust, and Contempt. 

To adapt this model for pain detection, we fine-tune the model using images from a pain dataset. We consider two datasets commonly used in automatic pain recognition - UNBC-McMaster shoulder pain expression database~\citep{c5} and BioVid heat pain dataset~\citep{walter2013biovid}. Models trained on these datasets have been shown to learn well-known facial expression patterns of pain~\citep{prajod2022using}. So, we use a combined dataset by merging UNBC and BioVid datasets. This increases the samples available for training a pain model. However, both these datasets are derived from video sequences and thus, have virtually repetitive images. To mitigate this redundancy, we select the images following the strategy proposed by~\cite{prajod2022using}. We also leverage the training, validation, and test dataset split that they proposed.

The prediction layer of the emotion recognition model is modified for a 2-class prediction of pain and no-pain classes. The entire network is fine-tuned using the combined pain dataset. Like in the case of attention detection, the images are face-cropped and scaled to the default VGG16 dimensions. The fine-tuning process employed an SGD optimizer (learning rate = 0.01) and focal loss function. The model achieved an average accuracy of 78\% on the test set.

\subsubsection{Stress Detection}
The patient may experience stress while performing the exercise, especially if they find the exercise hard or if they are unable to complete the recommended exercise. Such negative experiences can severely impact the user's level of motivation and their willingness to continue training. Hence, the user's stress level is an important input to the agent.

Unlike the other models, we rely on hand-crafted HRV (Heart Rate Variability) features to detect stress. This choice is based on the observations presented by~\cite{prajod2022generalizability}, where HRV features showed more generalizability than models based on raw ECG signals. We used the ECG signals from the WESAD dataset~\citep{c9} to derive the HRV features for training the stress detection model. This dataset contains physiological signals collected from 15 participants during a social stress scenario. We compute 22 HRV features from the time domain, frequency domain, and poincaré plots. These features are computed using 60-second-long ECG segments.  The pre-processing steps and feature extraction are detailed in~\cite{prajod2022generalizability}. 

We trained an SVM (Support Vector Machine) with the radial basis kernel function to predict if the user is stressed or not. To mitigate the individual differences in the signal and derived features (e.g. resting heart rate), the signals undergo MinMax normalization. This model achieves an average accuracy of 87\% in LOSO evaluation.

\subsubsection{Real-time Pipelines}

\begin{figure}
    \centering
    \includegraphics[width=0.99\textwidth]{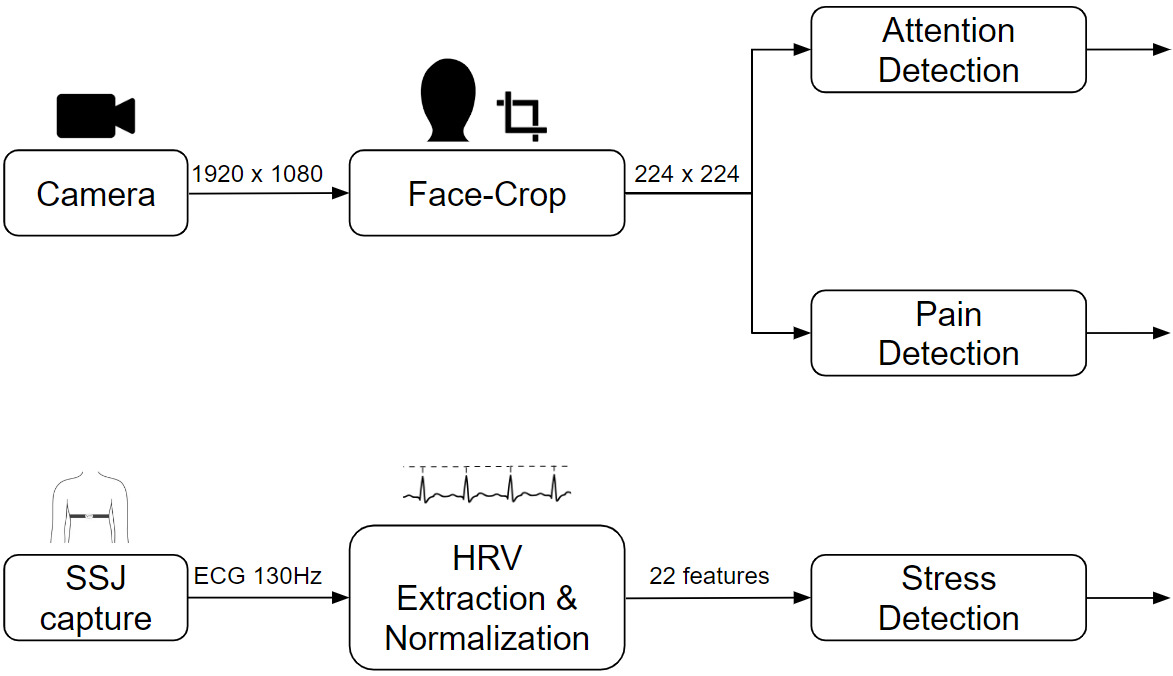}
    \caption{Illustration of pipelines deployed for real-time prediction of attention, pain, and stress.}
    \label{fig:pipes}
\end{figure}

To facilitate the real-time prediction of the user states, we employ two existing frameworks - SSI~\citep{WagnerSSI} and SSJ~\citep{10.3389/fict.2018.00013}. SSI is a Windows-based framework, whereas SSJ is developed for Android. The pipelines involving facial images including capture, processing, and predictions (attention, pain detection) are implemented using SSI. We use the SSJ plugins to capture the raw ECG signal from the Polar H10 device and stream it to SSI. SSI receives the raw ECG signals from SSJ and the subsequent pre-processing, feature extraction, and detection steps are performed within the SSI pipeline. Figure~\ref{fig:pipes} visualizes the SSI pipelines that were implemented and deployed.

We use a frontal face camera (Logitech RGB camera) to capture the facial expressions of the users. Each image from the video sequence is passed through a face crop plugin which crops the image to the face region and scales it to 224 $\times$ 224. The pre-processed image serves as input to the attention and pain detection models. The per-frame classification outputs from both models are communicated to the agent via UDP (User Datagram Protocol) sockets.

For stress detection, we use the Polar H10 chest band to collect ECG signals. We use the first 5 minutes to collect baseline data and compute the normalization parameters for each user. For the subsequent data, we compute the HRV features, normalize them, and detect stress using our SVM model. The prediction results are forwarded to the agent using UDP sockets.

\subsection{Interactive Agent}
\begin{figure}[h]
    \centering
    \includegraphics[width=0.6\textwidth]{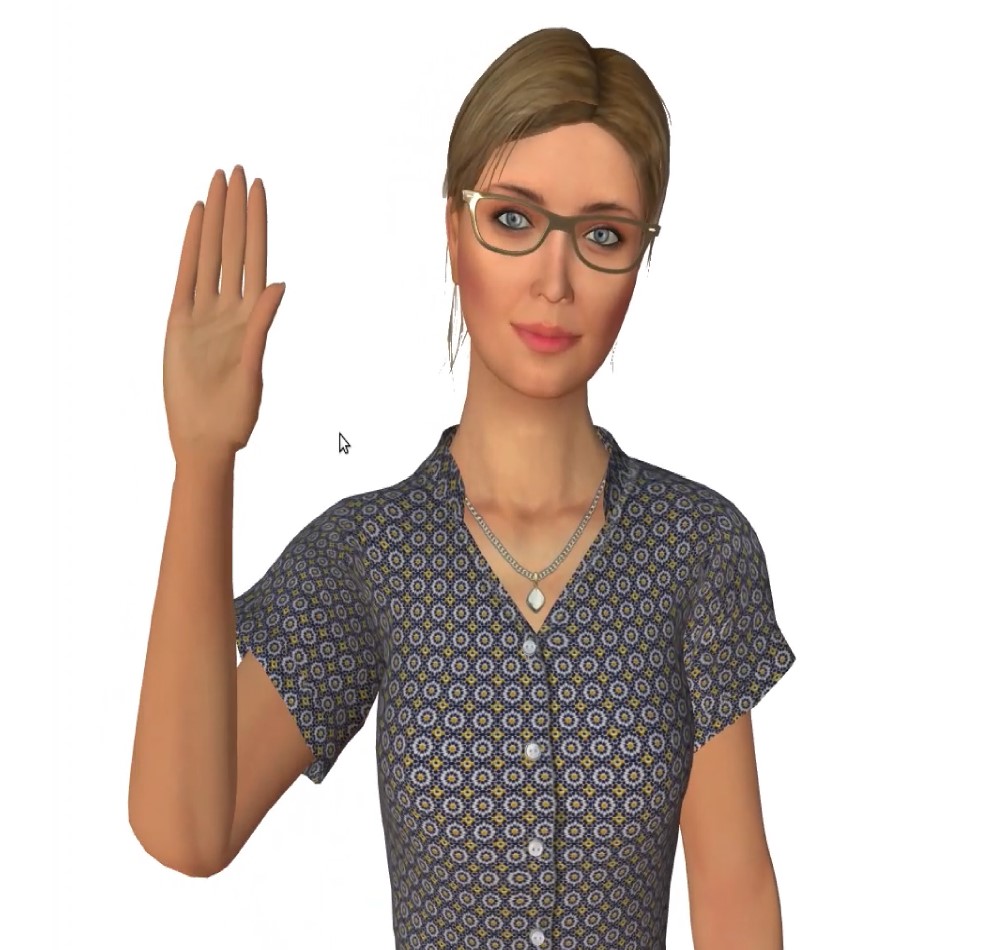}
    \caption{Socially Interactive Agent, Lydia.}
    \label{fig:lydia}
\end{figure}

A motivating agent is used in this paper to support the patients during their training sessions (Figure~\ref{fig:lydia}). The agent is displayed on a monitor along with the training exercises and instructions (refer to section~\ref{sec:gui}). The agent serves as a coach, motivating, informing, and assisting the patient with certain neurorehabilitation tasks. As stated in the introduction, it would be important that the agent's behavior and the rehabilitation device's behavior are calibrated in such a way that they appear to be one entity, with the agent assisting the patient in applying a particular amount of force through the physical capabilities made available by the robot. 

The agent has a human appearance and is developed by displaying narrow gestures near to the body, show facial expressions that can be related to positive emotions (e.g. joy, admiration, happy-for), using shorter pauses, and showing a friendly head and gaze behavior. On the verbal level, explanations and questions show appreciation for the user and contain many politeness phrases~\citep{Gebhard-et-al-14}.  The coaching assistant is designed to follow the best practices of training professionals (Section~\ref{sec:bestpractice}). Our approach includes the deliberate incorporation of human-like features and behavior in the virtual agent's design, thereby establishing an immediate connection with users. This anthropomorphic design choice goes beyond aesthetics; it serves as a conduit for users to attribute human-like motivations and intentions to the agent, reinforcing feelings of warmth and approachability. This parallels the significance of warmth and competence in human-human relationships and leverages the concept that individuals often apply the same social rules and expectations to virtual agents as they do to humans~\citep{nass2000machines,epley2007seeing}.

At the heart of our agent's design and implementation lies the fundamental concept of trust. Drawing inspiration from established principles of trust in human-human relationships, we have meticulously integrated key elements of warmth and competence into our virtual agent's behavior. To ensure that our agent embodies trustworthiness, we partnered with experienced psychologists and employed advanced tools like the VSM. Through iterative refinements, we fine-tuned the agent's verbal expressions to strike the right balance between warmth and competence. These interactions serve as a tangible representation for making the agent a trustworthy ally in the therapeutic journey.


\begin{figure}[h]
    \centering
    \includegraphics[width=0.99\textwidth]{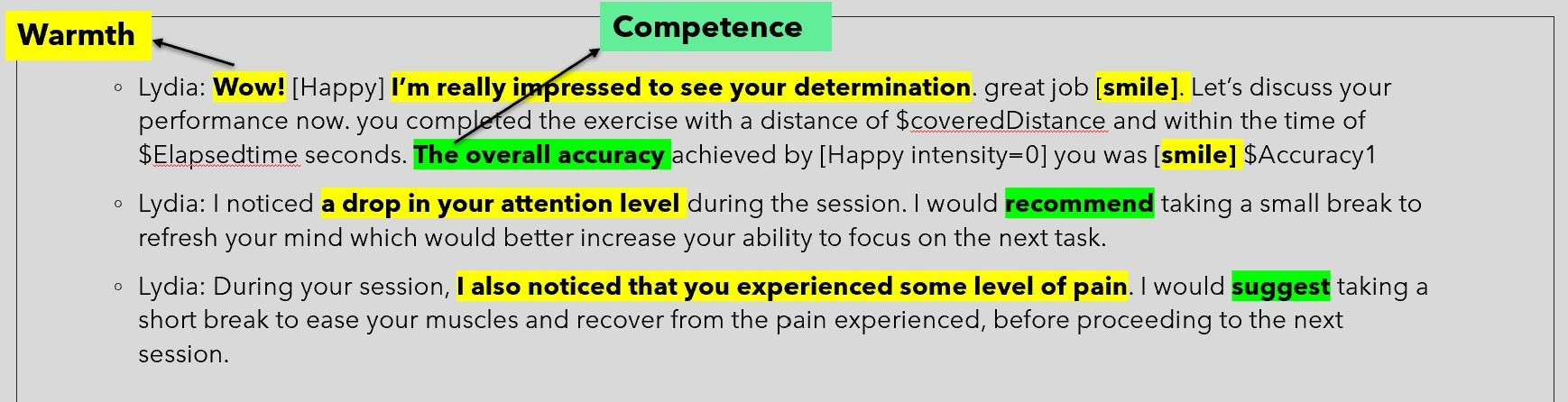}
    \caption{ Words and phrases that portrays the agent as warm and competent.}
    \label{fig:warmth}
\end{figure}

Our agent effectively balances warmth and competence in communication. Warmth is conveyed through expressions like "WOW!" and commendations such as "I’m impressed by your determination." In contrast, its competence is highlighted with terms like "overall accuracy," and action verbs like "suggest" and "recommend"~\citep{Gebhard-et-al-14}. The agent employs gestures, nods, and verbal affirmations, including smiles, to bolster user engagement and cultivate a conducive training atmosphere.

\begin{figure}[h]
    \centering
    \includegraphics[width=0.88\textwidth]{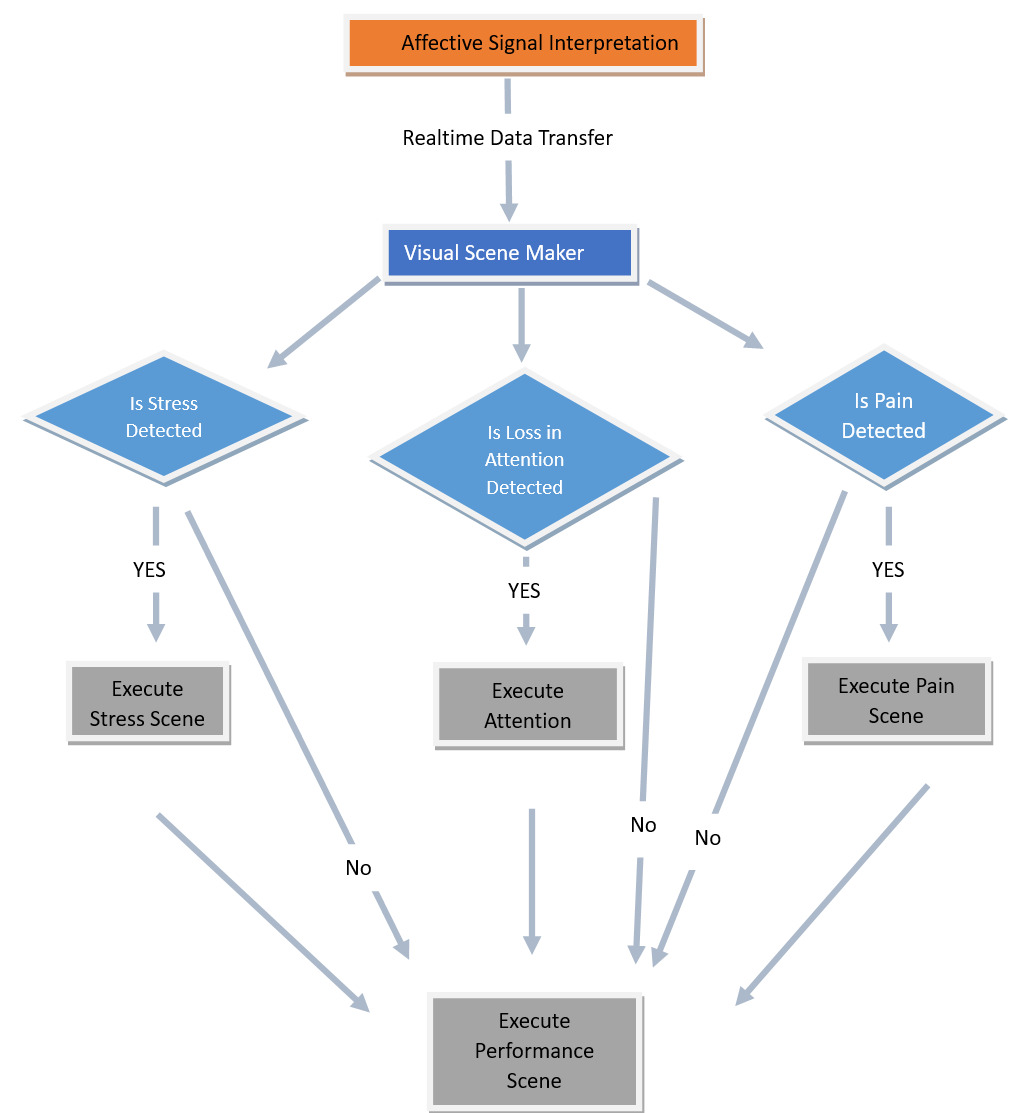}
    \caption{Flow chart showing how the affective signals are utilized in executing different scenarios in VSM }
    \label{fig:socialSignalsVSM}
\end{figure}

The agent's dynamism lies in its adaptability, seamlessly tailoring its behavior according to the user's physiological and affective cues. These cues, processed in real-time by a signal interpretation framework, enhance the agent's therapeutic relevance~\citep{charamel}. In the context of neurorehabilitation, metrics such as attention and pain are crucial. Hence, an empathic agent capable of identifying attention and pain contributes to the establishment of a rehabilitation environment that minimizes stress, proving essential for sustaining patient motivation. The interactive agent receives affective cues regarding stress, attention, and pain directly from the SSI pipeline into the VSM (Fig ~\ref{fig:socialSignalsVSM} ). If the value of any of these social cues exceeds its threshold, the agent is programmed to empathetically inform the user about their current state.

Hardware-wise, the interactive agent runs on a PC running MS Windows 10TM  and operates autonomously in a web browser, interacting via social cues~\citep{charamel}, showing the agent at a realistic size. It uses the CereProc Text-To-Speech system for voice outputs~\citep{cereproc}. Lydia, the agent, can execute 36 diverse conversational gestures~\citep{Schneeberger-et-al-21}. Additionally, 14 facial expressions, including Ekman's basic emotions, further its expressiveness~\citep{Ekman92}. All agent behaviors are streamlined through the VSM toolkit, ensuring dynamic and user-relevant content.

The agent engages with the system through a graphical user interface. Additionally, it proactively monitors users' affective signals (attention, pain, and stress) during the session. If, during the session, any of these social signals are detected, the agent empathically informs the user and suggests measures to mitigate the issue in the future (e.g., recommends taking a break). Following each session, Lydia delivers a comprehensive performance deviation index summary, subtly encouraging users to improve their future engagement(See figure~\ref{fig:socialSignalsVSM}, ).

\subsubsection{Ethical approval}
The study has been conducted according to the guidelines of the Declaration of Helsinki and approved by Commissione per l’Etica e l’Integrità nella Ricerca of the National Research Council of Italy (protocol n. 0085720/2022 of 23/11/2022). All 18 participants were briefed about the study and the details of data treatment before signing an informed consent from which they can withdraw at any point.

\section{Results}
\label{sec:results}

We conducted a small pilot study with 18 students, including 12 men and 6 women. The sample comprised 7 undergraduates and 11 master's students, with ages ranging from 22 to 33 years (M = 26.11, SD = 2.87, median = 25). The primary aim of the study was to evaluate our proposed framework. Specifically, we examined the effect of an interactive agent during therapy sessions on user engagement and gathered preliminary feedback on interaction quality for potential refinement.

\begin{figure}
    \centering
    \includegraphics[width=0.80\textwidth]{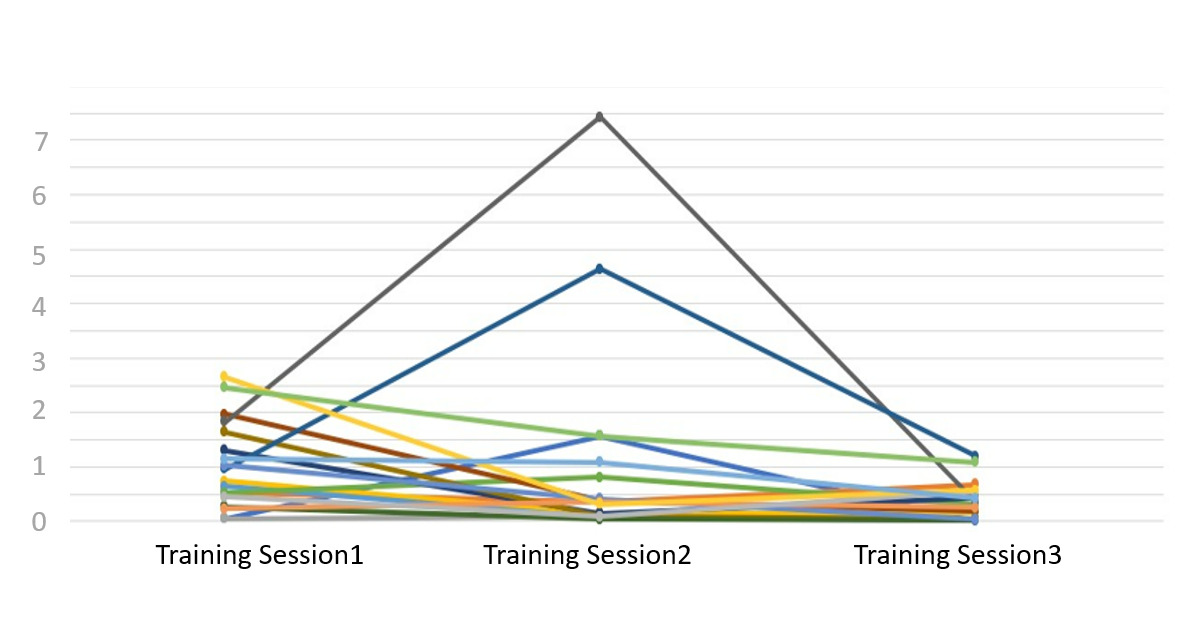}
    \caption{performance deviation index of all the participants in the training game.}
    \label{fig:trainingGame}
\end{figure}

Figure~\ref{fig:trainingGame} illustrates the performance deviation index of all participants in the training game. The y-axis represents the performance deviation index scores recorded during the training sessions (refer section~\ref{sec:gui}). The term "performance deviation index" is assessed based on three main criteria: (1) It is defined as the deviation of the actual path followed by the participant from the ideal path and is inversely proportional to the user's performance (2) the total distance traveled, and (3) the elapsed time. Figure~\ref{fig:trainingGame} specifically utilizes the performance deviation index to compare the performance of all participants. Excluding two notable outliers from Figure~\ref{fig:trainingGame},  there's a consistent downward trend in participants' overall performance deviation index throughout the study's duration. This suggests participants effectively adapted to the device, and the inclusion of the avatar did not negatively impact their performance. In other words, the avatar did not serve as a distraction to the participants during the training sessions. Additionally, the lower the performance deviation index, the better the performance. Therefore, in Figure~\ref{fig:trainingGame}, we see that the majority of the participants perform better in the third session, which means the performance deviation index is the lowest (i.e., close to zero) for the third session. In addition to the training game results, participants were administered a post-training questionnaire. The responses yielded valuable insights into the participants' experiences. One participant expressed a heightened sense of motivation, noting, \textit{"I felt more motivated to complete the task, knowing that someone was watching me."} This sentiment underscores the potential for the avatar's presence to positively influence user engagement and commitment.

Furthermore, another participant mentioned, \textit{"I felt very comfortable when interacting with the interactive agent ."} and there was just one user who had a blend of emotions, stating, \textit{"I felt a little intimidated at times and felt judged"}. While this response underscores the diversity of emotional reactions, it's worth noting that the majority of users positively appreciated the agent and its behavior. This sentiment aligns with the overall tone of participants' feedback, which leans overwhelmingly toward the positive side. Many participants reported experiencing heightened motivation and a heightened sense of accountability when interacting with the avatar. These collective observations are highly encouraging, suggesting that the avatar-assisted system has the potential to significantly boost user engagement and contribute positively to therapeutic outcomes, particularly in the context of neurorehabilitation.

\section{Discussion}
\label{sec:discussion}
The observed steady decline in participants' performance deviation index during the therapy sessions, as shown in Figure~\ref{fig:trainingGame}, underscores the potential effectiveness of our framework. This is particularly noteworthy given the novelty of introducing an interactive agent in such contexts. The data indicates a swift adaptation of participants to the device, and importantly, the presence of the interactive avatar did not act as a distraction, which counters some concerns previously raised in literature about interactive agents in therapeutic settings. However, it's imperative to approach these findings with a degree of caution, acknowledging the inherent limitations such as the sample size and study duration. These factors necessitate further extensive research to comprehensively understand the long-term impact and efficacy of such systems in rehabilitation outcomes. Our pilot study serves as a foundational step for more in-depth exploration in this domain. The encouraging results from this preliminary assessment underscore the importance of refining the interactive agent's features to enhance user engagement and exploring its applicability in various therapeutic contexts. As we continue to bridge the gap between technology and human-centric care, the integration of feedback from healthcare professionals will be pivotal in enhancing the system's efficacy and ensuring its alignment with clinical practices.

\section{Conclusions and Future works}
\label{sec:conclusions}
In conclusion, our innovative AI-based system stands as a transformative approach in the realm of neurorehabilitation, offering a viable solution to the scarcity of specialized care professionals. By harnessing the capabilities of a socially interactive agent integrated within a robotic framework, we have successfully demonstrated the potential to replicate the critical social interaction and motivation factors found in traditional therapy settings. The system's flexibility allows promoting at-home rehabilitation with less dependency on professional availability. Qualitative feedback from participants underscored the user interface and the virtual coach's anthropomorphic attributes as pivotal in maintaining engagement, with users reporting a heightened sense of companionship and support that spurred consistent use. The encouraging outcomes of our feasibility study with healthy patients, showcasing their adaptability to the system and heightened engagement without distraction, lay the groundwork for further research. Looking forward, we intend to expand the scope of our research by conducting extensive trials with real patients suffering from neuromotor dysfunctions. These future studies will not only allow us to validate the efficacy of our framework in a clinical setting but will also enable us to perform a comparative analysis of rehabilitation outcomes with and without the presence of the interactive agent. This will offer a clearer understanding of the agent's impact on patient engagement and recovery. Additionally, to ensure the robustness and generalizability of our findings, we plan to increase our sample size, providing a more comprehensive understanding of the system's effectiveness across a diverse patient demographic. Through these endeavors, we aspire to refine and validate our system, making a significant contribution to the field of neurorehabilitation and providing a path towards more accessible and personalized patient care.

\section*{Conflict of Interest Statement}
The authors declare that the research was conducted in the absence of any commercial or financial relationships that could be construed as a potential conflict of interest.

\section*{Author Contributions}
The authors confirm their contributions to the paper as follows. Study conception and design: Arora, R., Lavit Nicora, M. \& Prajod, P.; Data collection: Arora, R., Lavit Nicora, M. \& Tauro, G.; Analysis and interpretation of results: Arora, R. \& Prajod, P.; Draft manuscript preparation: Arora, R., Lavit Nicora, M., Prajod, P., Panzeri, D., \& Tauro, G.; Supervision and proof-reading: Vertechy, R., André, E., Malosio, M. \& Gebhard, P. All authors reviewed the results and approved the final version of the manuscript.

\section*{Funding}

The study was supported by the European Union’s Horizon 2020 research and innovation programme under the MindBot project (grant agreement number: 847926) and by the Bavarian Research Foundation under the FORSocialRobots project (reference number: 1594-23)

\section*{Acknowledgments}
This work was carried out within the framework of the AI Production Network Augsburg.



\bibliographystyle{unsrtnat} 
\bibliography{test}


\end{document}